SINTEF

# Report

## An Approach to Select Cost-Effective Risk Countermeasures Exemplified in CORAS

**Author(s)**
Le Minh Sang Tran
Bjørnar Solhaug
Ketil Stølen

# Report

# An Approach to Select Cost-Effective Risk Countermeasures Exemplified in CORAS

| | |
|---|---|
| **VERSION** | **DATE** |
| Final version | 2013-01-28 |

**AUTHOR(S)**
Le Minh Sang Tran
Bjørnar Solhaug
Ketil Stølen

| | |
|---|---|
| **CLIENT(S)** | **CLIENT'S REF.** |
| | |
| **PROJECT NO.** | **NUMBER OF PAGES/APPENDICES:** |
| | 17 + Appendices |

**ABSTRACT**
Risk is unavoidable in business and risk management is needed amongst others to set up good security policies. Once the risks are evaluated, the next step is to decide how they should be treated. This involves managers making decisions on proper countermeasures to be implemented to mitigate the risks. The countermeasure expenditure, together with its ability to mitigate risks, is factors that affect the selection. While many approaches have been proposed to perform risk analysis, there has been less focus on delivering the prescriptive and specific information that managers require to select cost-effective countermeasures. This paper proposes a generic approach to integrate the cost assessment into risk analysis to aid such decision making. The approach makes use of a risk model which has been annotated with potential countermeasures, estimates for their cost and effect. A calculus is then employed to reason about this model in order to support decision in terms of decision diagrams. We exemplify the instantiation of the generic approach in the CORAS method for security risk analysis.

| | |
|---|---|
| **PREPARED BY**<br>Le Minh Sang Tran | **SIGNATURE** |
| **CHECKED BY**<br>Quality Assuror | **SIGNATURE** |
| **APPROVED BY**<br>Project responsible | **SIGNATURE** |

| **REPORT NO.** | **ISBN** | **CLASSIFICATION** | **CLASSIFICATION THIS PAGE** |
|---|---|---|---|
| Report No. | ISBN | Unrestricted | Unrestricted |

# Table of Contents

# An Approach to Select Cost-Effective Risk Countermeasures Exemplified in CORAS

Le Minh Sang Tran[1], Bjørnar Solhaug[2], and Ketil Stølen[2]

[1] University of Trento, Italy
`tran@disi.unitn.it`
[2] SINTEF ICT, Norway
`{bjornar.solhaug,ketil.stolen}@sintef.no`

**Abstract.** Risk is unavoidable in business and risk management is needed amongst others to set up good security policies. Once the risks are evaluated, the next step is to decide how they should be treated. This involves managers making decisions on proper countermeasures to be implemented to mitigate the risks. The countermeasure expenditure, together with its ability to mitigate risks, is factors that affect the selection. While many approaches have been proposed to perform risk analysis, there has been less focus on delivering the prescriptive and specific information that managers require to select cost-effective countermeasures. This paper proposes a generic approach to integrate the cost assessment into risk analysis to aid such decision making. The approach makes use of a risk model which has been annotated with potential countermeasures, estimates for their cost and effect. A calculus is then employed to reason about this model in order to support decision in terms of decision diagrams. We exemplify the instantiation of the generic approach in the CORAS method for security risk analysis.

## 1 Introduction

In order to treat risks, decision makers (or managers) have to make decisions on proper countermeasures to be implemented to mitigate the risks. However, investment decisions are complicated. An organization needs the best possible information on risks and countermeasures to decide what is the best investment. This involves deciding which countermeasure offers a good trade-off between benefits and spending. The countermeasure expenditure, together with its ability to mitigate risks, are factors that affect the selection. Inappropriate and over-expensive countermeasures are money lost. Therefore, a systematic method that helps to reduce business exposure while balancing countermeasure investment against risks is needed. Such a method would thereby help answering questions like *"(1): How much is it appropriate to spend on countermeasures?"* and *"(2): Where should spending be directed?"* as highlighted by Birch and McEvoy [4].

Unfortunately, there exists little support for the prescriptive and specific information that managers require to select cost-effective risk countermeasures. Several cost estimation models have been proposed but most are only loosely coupled to risk analysis. For example, the Security Attribute Evaluation Method (SAEM) [7] is well-suited to evaluate risk reduction but is very vague on the issue of cost effectiveness. Likewise, [1] suggests several methods to assess cost of risks (*e.g.,* Cost-Of-Illness, Willingness-To-Pay), but none of these methods provides specific support to evaluate countermeasure expenditure. Chapman et al. [8] propose a framework which justifies mitigation strategies based on cost-difference but does not take the benefit-difference (*i.e.* level of risk reduction) between strategies into consideration.

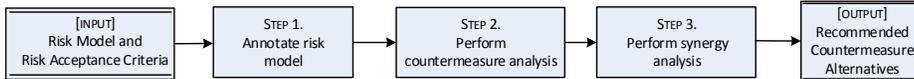

**Fig. 1.** Steps of our process.

Effective decision-making requires a correct risk model incorporating multi-aspect information on countermeasures and a method to select cost-effective countermeasure alternatives. The multi-aspect information should contain the knowledge about the countermeasures themselves, their expenditures and suitability to mitigate risks, as well as the impacts they can have on each other. The focus of this paper is not on how to obtain this information, but rather on how to make use of this information to select effective risk countermeasures. In particular, we propose a systematic approach to integrate such multi-aspect information to reason and make recommendations regarding cost-effective countermeasure alternatives. We are not aware of other approaches of this kind. Our approach is sufficiently generic to allow it to be combined with many existing risk analysis methods. In this paper, we demonstrate this by instantiating our generic approach in the CORAS method for security risk analysis [16] with concrete illustrative examples.

The structure of the paper is as follows. In Section 2 we describe our generic approach. Next, in Section 3 we exemplify the approach by instantiating it in the CORAS method. We present related work studies in Section 4. Finally, we summarise and draw conclusions in Section 5.

## 2 Our Approach

This section describes our approach aiming at the selection of cost-effective countermeasures for unacceptable risks. Section 2.1 provides an overview of this process and a conceptual model on which it builds. In Section 2.2 we describe the expectations to the risk model resulting from the risk assessment process and taken as input to our treatment assessment process. Finally, in Section 2.3 to Section 2.5 we describe its three main steps in further detail.

### 2.1 Process Overview

As illustrated in Fig. 1, our approach takes a risk model resulting from a risk assessment and the associated risk acceptance criteria as input and delivers a set of recommended countermeasure alternatives as output. Hence, the approach

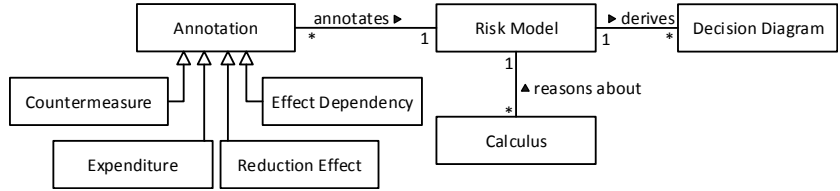

**Fig. 2.** Conceptual model.

assumes that risk assessment has already been conducted, i.e. that risks have been identified, estimated and evaluated and that the overall risk analysis process is ready to proceed with the risk treatment phase. We moreover assume that the risk analysis process complies with the ISO 31000 risk management standard [13], in which risk countermeasure is the final phase. Our process consists of three main steps as follows:

STEP 1 *Annotate risk model:* Identify and document countermeasures. The results are documented by annotating the risk model taken as input with relevant information including the countermeasures, their cost, their reduction effect (i.e., effect on risk value), as well as possible effect dependencies (i.e., countervailing effects among countermeasures).
STEP 2 *Perform countermeasure analysis:* Enumerate all countermeasure alternatives and reevaluate the risk picture for each alternative. The analysis makes use of the annotated risk model and a calculus for propagating and aggregating the reduction effect and effect dependency along the risk paths.
STEP 3 *Perform synergy analysis:* Perform synergy analysis for selected risks based on decision diagrams. The outcome is recommended countermeasure alternatives which cost-effectively mitigates the selected risks.

Fig. 2 presents the conceptual model, expressed as a UML class diagram [23] on which our approach builds. A *Risk Model* is a structured way of representing an unwanted incident and its causes and consequences using graphs, trees or block diagrams [22], or tables [16]. An unwanted incident is an event that harms or reduces the value of an asset, and a risk is the likelihood of an unwanted incident and its consequence for a specific asset [13]. A *Countermeasure* mitigates risk by reducing its likelihood and/or consequence. The *Expenditure* includes the expenditure of countermeasure implementation, maintenance and so on. The *Reduction Effect* captures the extent to which a countermeasure mitigates risks. The *Reduction Effect* could be the reduction of likelihood, and/or the reduction of consequence of a risk. The *Effect Dependency* captures the countervailing effect among countermeasures that must be taken into account in order to understand the combined effect of identified countermeasures. The *Calculus* provides a mechanism to reason about the annotated risk model. Using the *Calculus*, we can perform countermeasure analysis on annotated risk models to calculate the residual risk value for each individual risk. A *Decision Diagram* facilitates the decision making process based on the countermeasure analysis.

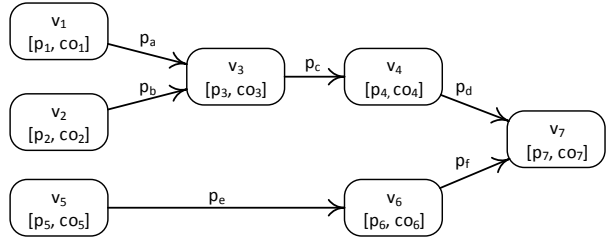

**Fig. 3.** Risk graph.

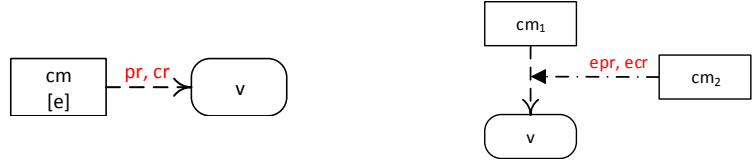

**Fig. 4.** Countermeasure with *treats* relation.    **Fig. 5.** Effect dependency relation.

### 2.2 Input Assumptions

The input of our approach is a risk model generated by a risk assessment, and risk acceptance criteria. To ensure that our approach is compatible with several risk modeling techniques, we expect the risk model could be understood as a risk graph instantiation. A risk graph [6] is a common abstraction of several established risk modeling techniques such as fault tree analysis (FTA) [11], event tree analysis (ETA) [12], attack trees [24], cause-consequence diagrams [17,22], Bayesian networks [9], and CORAS risk diagrams [16]. Hence, our approach complies with these risk modeling techniques, and can be instantiated by them.

A risk graph is a finite set of vertices and relations (see Fig. 3). Each vertex $v$ represents a threat scenario, *i.e.* a sequence of events that may lead to an unwanted incident, and can be assigned a probability $p$, and a consequence $co$. A *leads-to* relation from $v_1$ to $v_2$ means that the former threat scenario may lead to the latter. Probabilities on the relations are conditional probabilities indicating the likelihood of the former to lead to the latter when the former occurs.

### 2.3 Detailing of Step 1 – Annotate Risk Model

This step annotates the input risk model with required information for further analysis. There are four types of annotation as follows:

*Countermeasure:* In risk graphs, countermeasures are represented as rectangles. In Fig. 4 there is one countermeasure and this is named $cm$.

*Expenditure:* In risk graphs, expenditure is expressed within square brackets following the countermeasure name ($e$ in Fig. 4). This is an estimated of the total amount of money spent to ensure the mitigation of countermeasure including expenditure of implementation, deployment, maintenance, and so on.

*Reduction effect:* In risk graphs, reduction effect is represented by a dashed arrow decorated by two numbers (*pr* and *cr* in Fig. 4). It captures the mitigating effect of a countermeasure in terms of reduced likelihood (*i.e. probability reduction - pr*), reduced consequence (*i.e. consequence reduction - cr*), or both. Both *pr* and *cr* are relative percentage values, *i.e. pr, cr* $\in [0, 1]$.

*Effect dependency:* In risk graphs, effect dependency is represented by a dash-dot arrow with solid arrowhead decorated by two numbers (*effect on probability reduction (epr)*, and *effect on consequence reduction (ecr)* in Fig. 5). It captures the impact of a countermeasure to the reduction effect of another, *i.e.* it can increase or decrease *pr* and/or *cr* of another countermeasure. The *epr* impacts *pr* while the *ecr* impacts *cr*. Both *epr* and *ecr* are relative percentage values, *i.e. epr, ecr* $\in [0, 1]$.

### 2.4 Detailing of Step 2 – Countermeasure Analysis

The countermeasure analysis in this step is conducted for every individual risk of the annotated risk model. The analysis enumerates all possible countermeasure combinations, called *countermeasure alternatives* (or *alternatives* for short) and evaluates the residual risk value (*i.e.* residual consequence and probability) with resect to each alternative to determine the most efficient one. Residual risk value is obtained by propagating the reduction effect along the risk model to get the revised risk values. To this purpose, we have developed a calculus with propagation rules. An example of rule is shown as below.

**Rule 2.1** (Countermeasure)   If there is a treats relation from countermeasure *cm* to vertex $v(p, co)$ with probability reduction *pr* and consequence reduction *cr*, we have:

$$\frac{cm \xrightarrow{pr,cr} v \quad v(p, co)}{v(p \cdot \overline{pr}, co \cdot \overline{cr})}$$

Rule 2.1 applies to countermeasures as depicted in Fig. 4. The probability reduction *pr* on the probability *p* of the scenario means that *p* is reduced by $pr \in [0, 1]$. Hence, *p* is multiplied by $\overline{pr} = 1 - pr$. Likewise for the consequence reduction. The complete list of rules is presented in the Appendix B.

From the leftmost threat scenarios (*i.e.* scenarios that have only outgoing *leads-to* relations), probabilities assigned to threat scenarios are propagated to the right. During the propagation, probabilities assigned to *leads-to* relations and reduction effects of countermeasures are taken into account. Finally, the propagation stops at the rightmost threat scenarios (*i.e.* scenarios that have only incoming *leads-to* relations). Based on the results from the propagation, the residual risk value is computed.

*Decision Diagram* is a directed graph plotted in an X-Y plane for an individual risk as illustrated in Fig. 6. A node in a decision diagram is a *risk state*, which is a risk with a certain alternative applied. A *risk state* is represented by a triplet: probability, consequence, and countermeasure alternative. The first two values are obtained from the propagation on the risk diagram. A risk state is plotted as a node in the X-Y plane with the probability as the X coordinate, and the consequence as the Y coordinate. The countermeasure alternative is annotated on the path from the *initial state* $S_0$ where no countermeasure applied.

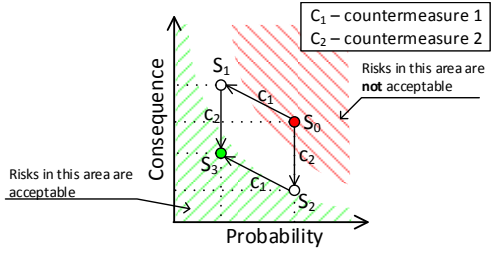

**Fig. 6.** Decision diagram.

To construct a decision diagram for a risk $v$, given $\mathcal{C}$ as set of countermeasures that can be implemented to mitigate $v$, we start at the initial state $S_0$. Then, from an existing state $S_i$, we reach new states by considering one or more further countermeasures in $\mathcal{C}$. By applying the analysis, we can calculate the corresponding risk state. Notice that we ignore all states whose residual consequence and probability are both greater than those of $S_0$ since it is useless to implement such countermeasures.

### 2.5 Detailing of Step 3 – Synergy Analysis

The aim of the synergy analysis is to recommend a cost-effective countermeasure alternative for mitigating all risks, namely *global countermeasure alternative*. Such recommendation is based on the decision diagrams for the individual risks (generated in Step 2), and the risk acceptance criteria, and the overall costs (OC) of global countermeasure alternatives which are calculated as follows:

$$\text{OC}(ca) = \sum_{r \in R_{ca}} \text{rc}(r) + \sum_{cm \in ca} \text{cost}(cm) \qquad (1)$$

where $ca$ is a global countermeasure alternative; $R_{ca}$ is the set of risks with respect to the global countermeasure alternative $ca$; rc() is a function that yields the loss (in monetary value) due to the risk taken as argument (based on its probability and consequence); cost() is a function that yields the expenditure of the countermeasure taken as argument.

The synergy analysis is decomposed into three following substeps:

STEP 3A *Identify global countermeasure alternatives:* Identify the set of global countermeasure alternatives $CA$ for which all risks are acceptable with respect to the risk acceptance criteria. Decision diagrams of individual risks can be exploited for identifying $CA$.

STEP 3B *Evaluate global countermeasure alternatives:* If no such global countermeasure alternative is identified ($CA = \emptyset$), do either of the following:
  - Identify new countermeasures and go to Step 1, or
  - Adjust the risk acceptance criteria and go to Step 3A

If some global countermeasure alternatives are identified ($CA \neq \emptyset$), select a global countermeasure alternative $ca \in CA$ with the lowest overall cost $OC(ca)$.

STEP 3C *Decide cost-effective global countermeasure alternative:* If $OC(ca)$ is acceptable (for the customer company in question) then terminate the analysis. Otherwise, identify more (cheaper and/or more effective) countermeasures and go to Step 1.

The above procedure may of course be detailed further based on various heuristics. For example, in many situations, with respect to Step 3A, if the global countermeasure alternative $ca \in CA$, then we do not have to consider other global countermeasure alternative $ca'$ such that $ca' \subseteq ca$. However, we do not go into these issues here.

## 3 Exemplification in CORAS

As a demonstration of applicability, this section instantiates the proposed approach into the CORAS method for security risk analysis [16] and exemplifies how the resulting extended CORAS method and language can be used to select cost-efficient risk countermeasures in an example drawn from a case study within the eHealth domain [21].

The *risk model* in the CORAS method is captured by so-called *risk diagrams*. A risk diagram is a causality graph consisting of potential causes (*i.e. threats*) that might (or might not) exploit flaws, weaknesses, or deficiencies (*i.e. vulnerabilities*) causing a series of events (*i.e. threat scenarios*) to happen, which could lead to *unwanted incidents* with certain likelihood and concrete consequence (*i.e. risks*) to a particular *asset*. Threat scenarios and risks are also called core elements in the risk diagram notation.

In the risk diagram, there are two kinds of relationships with assigned likelihoods: *initiate* and *leads-to* relations. The former connects a threat to a core element, and the latter connects a core element to another core element. Likelihoods assigned to *initiate* relations can be either *probabilities* or *frequencies*, whereas, likelihoods assigned to *leads-to* relations are *conditional likelihoods*.

Any risk diagram can be understood as an instantiation of a risk graph; such conversion is formally defined in [6]. In this paper, to save space we adjust the steps of the generic method such that they work directly with CORAS artifacts. To make the instantiation more comprehensible, we also present a running example that exploits an eHealth scenario proposed by the NESSoS project [19] to exemplify the resulting extended CORAS method.

### 3.1 eHealth Running Example: Patient Monitoring

As illustrated in Fig. 7, patients' behaviors and symptoms are monitored in realtime. This provides an improved basis for disease diagnoses and tailored therapy prescription regiments. Patients are equipped with sensors that continuously collect patients data and send these data to a handheld smart device (*e.g.,* smart

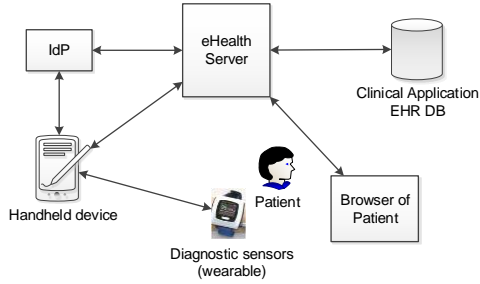

The patient has one or more monitoring devices in the form of wearable sensors. They provide data to an application in a handheld device, which does some processing on the data, aggregates the results and sends them to the eHealth Server. The patient and his devices are authenticated by an Identity Provider (IdP).

**Fig. 7.** Architectural sketch of Patient Monitoring scenario (from [19, Figure 3.2])

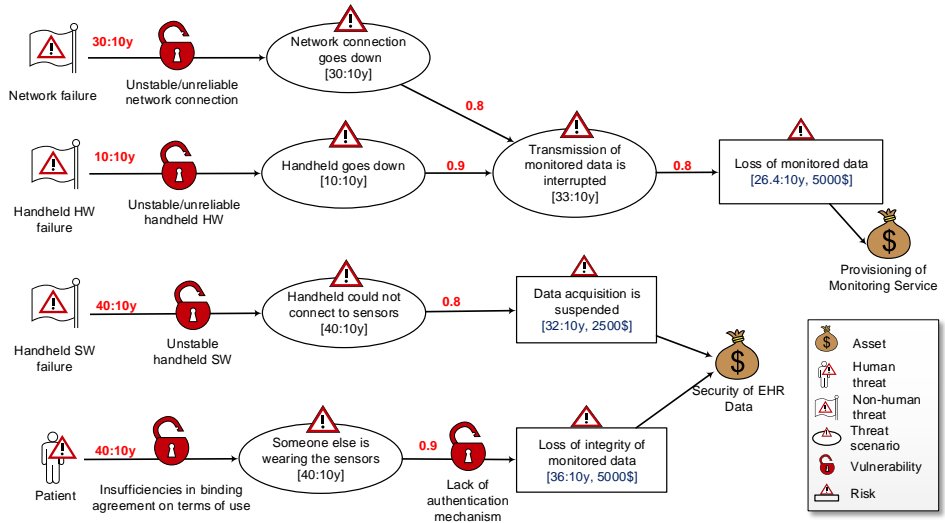

**Fig. 8.** Risk diagram of the scenario.

phone). This smart device, in turn, sends the patient data to external eHealth servers where the patients' eHealth Records (EHRs) are updated.

The CORAS risk diagram in Fig. 8 presents a partial result from a risk analysis of the Patient Monitoring scenario [21]. In this risk diagram, *network failure* exploits the vulnerability *unstable/unreliable network connection* to initiate *network connection goes down*. Likewise, *handheld HW failure* exploits the vulnerability *unstable/unreliable handheld HW* to initiate *handheld goes down*. Both *handheld goes down* as well as *network connection goes down* may lead to the *transmission of monitored data is interrupted*. This, consequently, may lead to *loss of monitored data* which impacts the *provisioning of monitoring service*. The rest of the diagram is interpreted in the similar manner.

We assume in the following that this diagram is a consistent and complete documentation of risks identified during the risk assessment. We moreover use frequencies to estimate likelihoods of core elements. Reasoning about frequencies in the risk and treatment assessment is also supported by our calculus.

### 3.2 Applying Step 1 – Annotate Risk Model

In this step, we annotate the CORAS risk diagrams according to Step 1 to create CORAS treatment diagrams. Note that in CORAS, countermeasures are referred to as treatments.

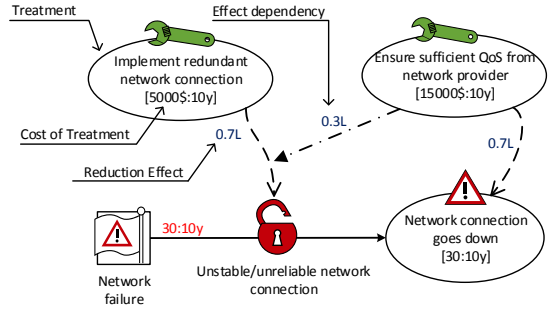

**Fig. 9.** Annotated diagram

*Treatment annotation:* treatments can apply to most of the elements in a treatment diagram, including all types of core elements, threats, and vulnerabilities. Fig. 9 shows an example in which a treatment *implement redundant network connection* treats the scenario *network connection goes down* which was initiated by *network failure* by exploiting the vulnerability *unstable/unreliable network connection*.

*Expenditure annotation:* the treatment expenditure, annotated as value inside the treatment bubble, is the total expenditure spent for a treatment in a period of time. For instance, in Fig. 9, the expenditure for *implementing a redundant network connection* is 5000USD:10$y$ in ten year.

*Reduction effect annotation:* following Step 1, reduction effect in the CORAS instantiation is annotated on *treats* relations as a pair $\langle fr, cr \rangle$, where $fr, cr$ are frequency reduction and consequence reduction, respectively. For example, in Fig. 9, the frequency of network failure is thirty times in ten years, annotated as 30:10$y$. In a CORAS diagram, we suffix the value of $fr$ and $cr$ with the letter 'L' and 'C', respectively, to distinguish between them. The treatment *implement redundant network connection* only reduces the frequency (not consequence) of *unreliable network connection* by 0.7 at cost 5000USD:10$y$. This means the reduced frequency is $(1 - 0.7) \cdot 30{:}10y = 9{:}10y$.

*Effect dependency annotation:* in Fig. 9, to mitigate *network connection goes down*, we could *ensure sufficient Quality-of-Service (QoS) from network provider* with the cost of 15000USD:10$y$. This, however, reduces the effect of a redundant connection. These two treatments are countervailing. Ensuring such QoS will reduce the reduction effect of a redundant connection by 0.3 as annotated in the figure.

As summary, Fig. 10 shows the treatment diagram resulting from annotating the risk diagram in Fig. 8. Note that the likelihood annotations in Fig. 10 are after the application of the analysis of Step 2, which is explained next.

### 3.3 Applying Step 2 – Treatment Analysis

The analysis employs an instantiated version of the calculus for risk graphs. Here, we exemplify the propagation of likelihoods and reductions through an example taken from the annotated treatment diagram of the eHealth scenario. Particularly, we show how to do the propagation for risk *"Loss of Monitored Data"* (LMD). The result is presented in Fig. 10. For clarity, we use following acronyms for text in the diagram:

- TDI: *"Transmission of monitored Data is Interrupted"*
- NCD: *"Network Connection goes Down"*
- HGD: *"Handheld Goes Down"*
- NF: *"Network Failure"*
- HHW: *"Handheld HW failure"*
- IRN: *"Implement Redundant Network connection"*
- EQS: *"Ensure sufficient QoS from network provider"*
- IRH: *"Implement Redundant Handheld"*

Here we describe the frequency propagated for LMD. First, NF initiates NCD with frequency $30{:}10y$. The treatment IRN would reduce this frequency by $0.7L$. However, due to the effect dependency of EQS to IRN, the likelihood reduction of IRN is changed to $0.7L \cdot (1 - 0.3L) \approx 0.5L$. Hence, IRN reduces the frequency propagated to NCD to $30{:}10y \cdot (1 - 0.5) = 15{:}10y$. EQS would reduce the frequency of NCD by $0.7L$. So, the frequency propagated to NCD is $15{:}10y \cdot (1 - 0.7) = 4.5{:}10y$. Second, HHW initiates HGD with frequency $10{:}10y$. This is propagated to HGD. IRH treats HGD with likelihood reduction $0.7L$. Hence, frequency propagated to HGD is $10{:}10y \cdot (1 - 0.7) = 3{:}10y$. Since NCD and HGD are independent and both of them *lead-to* TDI, the frequency propagated to TDI is $4.5{:}10y \cdot 0.8 + 3{:}10y \cdot 0.9 = 6.3{:}10y$. Finally, the propagated frequency of LMD is $6.3{:}10y \cdot 0.8 = 5.04{:}10y$.

Likewise we can calculate the frequencies of the entire diagram. Fig. 10 shows the complete diagram with frequencies calculated and annotated. Note that in Fig. 10 we have calculated the likelihoods when all treatments are taken into account. However, due to the effect dependencies it may be that implementing all treatments is not the optimal alternative.

Fig. 11 presents decision diagrams of risk LMD and LID (*i.e. Loss of Integrity of monitored Data*). The detail result of the countermeasure analysis for risk LMD is provided in the Appendix A.

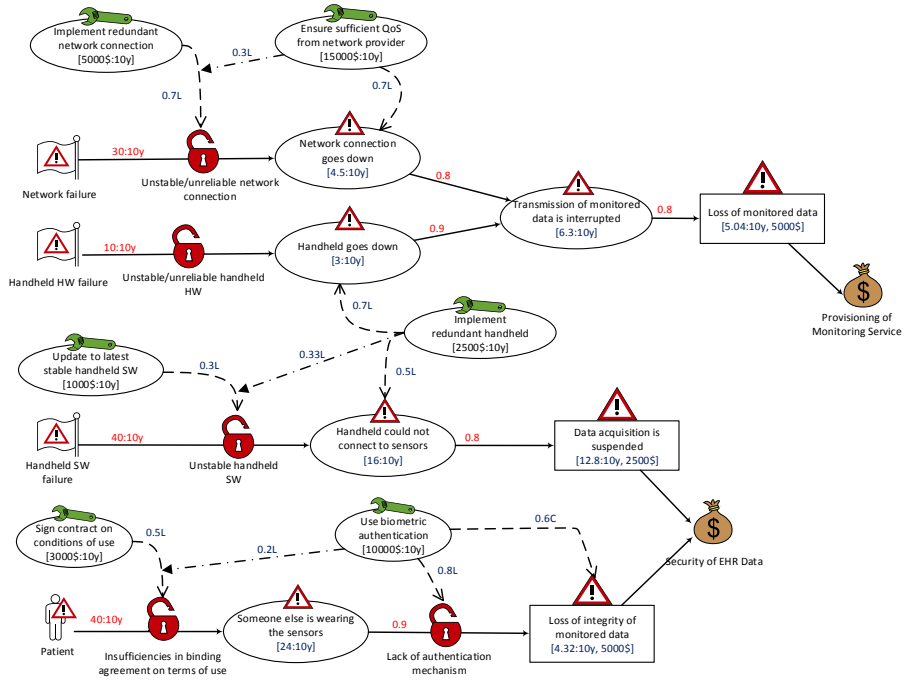

**Fig. 10.** Annotated treatment diagram with frequencies propagated.

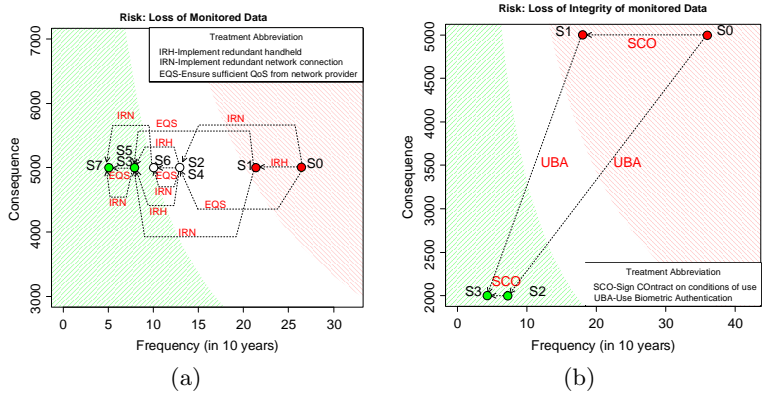

**Fig. 11.** Decision diagrams of risks in the eHealth scenario.

### 3.4 Applying Step 3 – Synergy Analysis

To facilitate the synergy analysis described in Step 3, we define the rc() function in (1) as follows: $rc(r) = co \cdot f$, where $co$ is the consequence and $f$ is the frequency of the risk $r$. Having decision diagrams for individual risks, the synergy analysis described in Step 3 is detailed as below.

STEP 3A We identify the set of global treatment alternatives based on the decision diagrams generated in the previous step and the expenditures of treatments. For each risk, we select the alleged cost-effective treatment alternatives. In particular, we choose $S3_{LMD}\{IRH, IRN\}$ (*i.e.* state $S3$ of risk $LMD$), and $S7_{LMD}\{IRH, IRN, EQS\}$; for risk DAS, we choose $S3_{DAS}\{USW, IRH\}$, and $S2_{DAS}\{IRH\}$; for risk LID, we choose $S3_{LID}\{IRH, IRN\}$ and $S2_{LID}\{UBA\}$. We assume all of these alternatives are acceptable with respect to the acceptance criteria.

STEP 3B We calculate the overall cost for each global treatment alternative using (1). Table 1 reports the overall costs for these alternatives. According to this table, we select GS1 due to its smallest overall cost.

STEP 3C For the sake of simplicity, we assume that customers are satisfied with the recommendation. Therefore, GS1 will be chosen for implementation.

Table 1. The global treatment alternatives in synergy analysis.

| Global Treatment Alternative | Individual Risk | | | Overall Cost |
|---|---|---|---|---|
| | LID | LMD | DAS | |
| GS1{UBA,SCO,IRH,IRN,USW} | S3 | S3 | S3 | 101740 |
| GS2{UBA,SCO,IRH,IRN,EQS,USW} | S3 | S7 | S3 | 102340 |
| GS3{UBA,IRH,IRN,USW} | S2 | S3 | S3 | 104500 |
| GS4{UBA,IRH,IRN,EQS,USW} | S2 | S7 | S3 | 105100 |
| GS5{UBA,SCO,IRH,IRN} | S3 | S3 | S2 | 108740 |
| GS6{UBA,SCO,IRH,IRN,EQS} | S3 | S7 | S2 | 109340 |
| GS7{UBA,IRH,IRN} | S2 | S3 | S2 | 111500 |
| GS8{UBA,IRH,IRN,EQS} | S2 | S7 | S2 | 112100 |

## 4 Related Work

Mehr and Forbes [18] suggest that "risk management theory needs to merge with traditional financial theory in order to bring added realism to the decision-making process". In line with the suggestion, Cost-benefit analysis (CBA) is often used with risk management to assess the effectiveness of risk countermeasures [2, 5, 25]. Major CBA steps include: *a)* develop measures to mitigate a certain problem *b)* develop measure alternatives *c)* estimate the impact and cost of each measure *d)* compare the benefit and costs for each measure alternative *e)* conduct a sensitive analysis of the uncertainty of estimated benefit and cost *f)* recommend a cost-effective measure alternative for implementation. Our approach may be seen as a special case or refinement of this process.

In risk management, decision on different risk mitigation alternatives has been emphasized in many studies [1, 20, 26]. The guideline in [26] proposes cost-benefit analysis to optimally allocate resources and implement cost-effective controls after identifying all possible countermeasures. This encompasses the determination of the impact of implementing (and not implementing) the mitigations, and the estimated costs of them. Another guideline [1] provides a semi-quantitative risk assessment. The probability, impact of risks are put into categories which are assigned with scores. The differences between the total score for all risks before and after any proposed risk reduction strategy relatively show the efficiency among strategies, and effectiveness of their costs. It also suggests that the economic costs for baseline risks should be evaluated using one of the following methods: Cost-Of-Illness, Willingness-To-Pay, Qualified-Adjusted Life Years, Disability-Adjusted Life Years. However, these methods have not been designed to assess cost of treatments but rather cost of risks.

Norman [20] advocates the use of Decision Matrix to agree on countermeasure alternative. A Decision Matrix is a simple spreadsheet which contains a list of countermeasures and a list of risks which those countermeasures mitigate. For each countermeasure there are estimates with respect to cost, effectiveness, and convenience. The countermeasure effectiveness is measured by metrics contained within the Sandia Vulnerability Assessment Model. The proposed approach is however not clearly defined, and all metrics are developed as spreadsheets which are complicated to implement and follow. Meanwhile, our proposal is graphical and backed up with a formal definition and reasoning. Butler [7] proposes the Security Attribute Evaluation Method (SAEM) to evaluate alternative security designs. It employs a four-step process, namely benefit assessment, threat index evaluation, coverage assessment, and cost analysis. This approach however focuses mostly on the consequence of risks rather than cost of countermeasures.

Chapman and Leng [8] describes a decision methodology to measure the economic performance of risk mitigation alternatives. The methodology is based on two kinds of analysis (baseline and sensitivity), four methods of economic evaluation, and a cost-accounting framework. The cost is broken down into several dimensions and types. The advantage is to provide a clear economic justification among mitigation alternatives. However, it does not differentiate alternatives based on their suitability to mitigate risks. In other words, the methodology focuses on the cost-difference aspect but does not take into account the benefit-difference (in terms of level of risks reduced) among alternatives.

Houmb et al. [10] introduce SecInvest, a security investment support framework which derives a security solution fitness score to compare alternatives and decide whether to invest or to take the associated risk. SecInvest relies on an eight-step trade-off analysis which employs existing risk assessment techniques for risk level. SecInvest scores alternatives with respect to their cost and effect, trade-off parameters, and investment opportunities. However, this approach does not provide a systematic way to assess the effects of alternatives on risks, either not take into account the dependency among countermeasures in an alternative.

There exist studies on Real Options Thinking [3, 14, 15] to articulate and compare different security solutions in terms of their business value. These solutions however are on the management aspect such as postpone, abandon, or continue to invest in security. Meanwhile, our alternatives are more focused on the technical aspect. The output of our approach could be taken as the input for Real Options Thinking based assessment.

## 5 Conclusion

We have presented a generic approach to select a cost-effective countermeasure alternative to mitigate risks. The approach requires input in the form of risk models represented as risk graphs. The approach analyses risk countermeasures with respect to different properties such as the amount of risk mitigation (Reduction Effect), how countermeasures affect others (Effect Dependency), and how much countermeasures cost (Countermeasure Expenditure). We have developed a set of formal rules extending the existing calculus for risk graphs. These new rules propagate the likelihoods and consequences along risk graphs thereby facilitating a quantitative countermeasure analysis on individual risks, and a synergy analysis on all the risks. The outcome is a list of countermeasure alternatives quantitatively ranked. These alternatives are represented not only in tabular format, but also in graphical style (Decision Diagram).

We have exemplified the generic approach by embedding it within the CORAS method. We extend the CORAS method with our approach in an example of the eHealth domain to select cost-effective treatments. Notations and rules have been adapted to comply with CORAS. The example demonstrates that our approach can work with existing defensive risk analysis methods whose risk models can be converted to risk graphs.

## Acknowledgement

This work is supported by the European Commission under project EU-IP-NESSoS. We would like to thank colleagues in the FRISK project at the SINTEF ICT, who have contributed substantial materials and suggestions for the risk analysis of the eHealth scenario.

## A  Analysis for Risk: Lost of integrity of Monitored Data

## B  Rules for Reasoning About Frequencies in CORAS

**Table 2.** Analysis for the risk LMD.

The name of each treatment alternative is shown in the first column (Risk State). The *Frequency* column is number of occurrences in ten years. Both *Frequency* and *Consequence* columns are values after considering the treatments.

| Ensure sufficient QoS from network provider |
| Implement Redundant Network connection |
| Implement Redundant Handheld |

| Risk/Risk State | Treatment | Frequency | Consequence |
|---|---|---|---|
| *Risk : Loss of Monitored Data* | | | |
| S0 |  | 26.4 | 5000 |
| S1 | ● | 21.36 | 5000 |
| S2 | ● (col 2) | 12.96 | 5000 |
| S3 | ● ● | 7.92 | 5000 |
| S4 | ● (col 3) | 12.96 | 5000 |
| S5 | ● ● (col 1,3) | 7.92 | 5000 |
| S6 | ● ● (col 2,3) | 10.08 | 5000 |
| S7 | ● ● ● | 5.04 | 5000 |

# Rules for Reasoning About Frequencies and Consequences in CORAS

Ketil Stølen

February 26, 2013

**Abstract**

# Contents

# 1 Formal foundation

In the following we introduce the formal machinery.

## 1.1 Basics

$\mathbb{N}$ and $\mathbb{R}$ denote the sets of natural numbers and real numbers, respectively. We use $\mathbb{N}_0$ to denote the set of natural numbers including 0, while $\mathbb{R}^+$ denotes the set of nonnegative real numbers. This means that:

$$\mathbb{N}_0 \stackrel{\text{def}}{=} \mathbb{N} \cup \{0\}, \quad \mathbb{R}^+ \stackrel{\text{def}}{=} \{r \in \mathbb{R} \mid r \geq 0\}$$

For any set of elements, we use $\mathbb{P}(A)$ to denote the powerset of $A$.

A tuple is an element of a Cartesian product. We use $\pi_j$ to extract the $j$th element of a tuple. Hence, if

$$(a, a') \in A \times A'$$

then $\pi_1.(a, a') = a$ and $\pi_2.(a, a') = a'$.

## 1.2 Sequences

By $A^\infty$, $A^\omega$ and $A^*$ we denote the set of all infinite sequences, the set of all finite and infinite sequences and the set of all finite sequences over some set of elements $A$, respectively. Hence, we have that

$$A^\omega = A^\infty \cup A^*$$

We define the functions

$$\#\_ \in A^\omega \to \mathbb{N}_0 \cup \{\infty\}, \quad \_[\_] \in A^\omega \times \mathbb{N} \to A$$

to yield the length and the $n$th element of a sequence. Hence, $\#s$ yields the number of elements in $s$, and $s[n]$ yields the $n$th element of $s$ if $n \leq \#s$.

We also need functions for concatenation and filtering:

$$\_\frown\_ \in A^\omega \times A^\omega \to A^\omega, \quad \_\circledS\_ \in \mathbb{P}(A) \times A^\omega \to A^\omega$$

Concatenating two sequences implies gluing them together. Hence, $s_1 \frown s_2$ denotes a sequence of length $\#s_1 + \#s_2$ that equals $s_1$ if $s_1$ is infinite, and is prefixed by $s_1$ and suffixed by $s_2$, otherwise.

The filtering operator is used to filter away elements. $B \circledS s$ denotes the subsequence obtained from $s$ by removing all elements in $s$ that are not in the set $B$.

## 1.3 Timed events

$\mathbb{E}$ denotes the set of all events, while the set of all timestamps is defined by

$$\mathbb{T} \stackrel{\text{def}}{=} \mathbb{R}^+$$

A timed event is an element of

$$\mathbb{E} \times \mathbb{T}$$

## 1.4 Histories

A history is an infinite sequence of timed events that is ordered by time and progresses beyond any finite point in time. Hence, a history is an element of

$$\mathbb{H} \stackrel{\text{def}}{=} \{ \quad h \in (\mathbb{E} \times \mathbb{T})^\infty \mid$$
$$\forall n \in \mathbb{N} : \pi_2.h[n] \leq \pi_2.h[n+1]$$
$$\forall t \in \mathbb{T} : \exists n \in \mathbb{N} : \pi_2.h[n] > t \quad \}$$

The first conjunct requires the timestamp of a timed event to be at least as great as that of its predecessor. The second conjunct makes sure that time will always progress beyond any finite point in time. That is, for any timestamp $t$ and history $h$ there is a timed event in $h$ whose timestamp is greater than $t$.

We also need a function for truncating histories

$$\_|\_ \in \mathbb{H} \times \mathbb{T} \to (\mathbb{E} \times \mathbb{T})^*$$

The truncation operator captures the prefix of a history upto and including a certain point in time. Hence, $h|_t$ describes the maximal prefix of $h$ whose timed events all have timestamps less than or equal to $t$.

## 1.5 Frequencies

As explained above, we use the nonnegative real numbers to represent time. The time unit is equal to 1. For simplicity, we assume that all frequencies are per time unit. The set of frequencies $F$ is therefore defined as follows:

$$\mathbb{F} \stackrel{\text{def}}{=} \mathbb{R}^+$$

Hence, $f \in \mathbb{F}$ denotes the frequency of $f$ occurrences per time unit.

# 2 Risk graphs

## 2.1 Syntax of risk graph formulas

### 2.1.1 Risk graphs

A risk graph is a pair of two sets $(V, R)$ where

$$V \subseteq \mathbb{P}(\mathbb{E}) \times \mathbb{F}, \quad R \subseteq V \times \mathbb{R}^+ \times V$$

We refer to the elements of $V$ as vertices and to the elements of $R$ as relations. We use $v(f)$ to denote a vertex, while $v \xrightarrow{r} v'$ denotes a relation.

### 2.1.2 Vertex expressions

The set of vertex expressions is the smallest set $X_V$ such that

$$\mathbb{P}(\mathbb{E}) \subseteq X_V, \quad v, v' \in X_V \Rightarrow v \sqcup v' \in X_V \wedge v \sqcap\!\!| \, v' \in X_V$$

We need a function

$$s \in X_V \to \mathbb{P}(\mathbb{E})$$

that for any vertex expression yields its set of events. Formally, $s$ is defined recursively as follows:

$$s(v) \stackrel{\text{def}}{=} \begin{cases} v & \text{if } v \in \mathbb{P}(\mathbb{E}) \\ s(v_1) \cup s(v_2) & \text{if } v = v_1 \sqcup v_2 \\ s(v_2) & \text{if } v = v_1 \sqcap\!\!| \, v_2 \end{cases}$$

### 2.1.3 Risk graph formula

A risk graph formula is of one of the following two forms

$$H \vdash v(f), \quad H \vdash v \xrightarrow{r} v'$$

where

- $H \in \mathbb{P}(\mathbb{H}) \setminus \varnothing$,
- $v, v' \in X_V$,
- $f \in \mathbb{F}$,
- $r \in \mathbb{R}^+$.

## 2.2 Semantics of risk graph formulas

We use the brackets $[\![\ ]\!]$ to extract the semantics of a risk graph formula. If $v \in \mathbb{P}(\mathbb{E})$ we define:

$[\![\ H \vdash v(f)\ ]\!] \stackrel{\text{def}}{=}$
  $\forall\, h \in H :$
    $f = \lim_{t \to \infty} \frac{\#((v \times \mathbb{T})\ \text{\textcircled{S}}\ (h|_t))}{t}$

The semantics of any other risk graph formula is defined recursively as follows:

$[\![\ H \vdash v_1 \sqcup v_2(f)\ ]\!] \stackrel{\text{def}}{=}$
  $\exists\, f_1, f_2, f_3 \in \mathbb{F} :$
    $[\![\ H \vdash v_1(f_1)\ ]\!]$
    $[\![\ H \vdash v_2(f_2)\ ]\!]$
    $[\![\ H \vdash s(v_1) \cap s(v_2)(f_3)\ ]\!]$
    $f_1 + f_2 - f_3 \leq f \leq f_1 + f_2$

$[\![\ H \vdash v_1 \sqcap v_2(f)\ ]\!] \stackrel{\text{def}}{=}$
  $\exists\, r \in \mathbb{R}^+;\ f_1, f_2 \in \mathbb{F} :$
    $[\![\ H \vdash v_1(f_1)\ ]\!]$
    $[\![\ H \vdash v_2(f_2)\ ]\!]$
    $f = f_1 \cdot r$
    $f \leq f_2$

$[\![\ H \vdash v_1 \xrightarrow{r} v_2\ ]\!] \stackrel{\text{def}}{=}$
  $\exists\, f_1, f_2 \in \mathbb{F} :$
    $[\![\ H \vdash v_1(f_1)\ ]\!]$
    $[\![\ H \vdash v_2(f_2)\ ]\!]$
    $f_2 \geq f_1 \cdot r$

## 2.3 Calculus of risk graph formulas

The three rules below correspond to rules 13.10, 13.11 and 13.12 in the CORAS book, respectively. There are some minor differences. In the CORAS book the real number decorating a leads-to relation is restricted to $[0, 1]$. The statistical independence constraint in Rule 13.12 of the CORAS book is not needed.

### 2.3.1 Rule for leads-to

$$\frac{H \vdash v_1(f) \quad H \vdash v_1 \xrightarrow{r} v_2}{H \vdash v_1 \sqcap v_2(f \cdot r)}$$

**Soundness**  Assume

(1)  $H \vdash v_1(f)$

(2)  $H \vdash v_1 \xrightarrow{r} v_2$

Then

(3)  $H \vdash (v_1 \sqcap v_2)(f \cdot r)$

Proof: (2) implies there are $f_1, f_2 \in \mathbb{F}$ such that

(4)  $[\![ H \vdash v_1(f_1) ]\!]$

(5)  $[\![ H \vdash v_2(f_2) ]\!]$

(6)  $f_2 \geq f_1 \cdot r$

(1) and (4) imply

(7)  $f = f_1$

(6) and (7) imply

(8)  $f_2 \geq f \cdot r$

(4), (5) and (8) imply (3).

### 2.3.2  Rule for mutually exclusive vertices

$$\frac{H_1 \vdash v_1(f) \wedge v_2(0) \quad H_2 \vdash v_2(f) \wedge v_1(0)}{H_1 \cup H_2 \vdash v_1 \sqcup v_2(f)}$$

For simplicity we have merged four premises into two using logical conjunction.

**Soundness**  Assume

(1)  $H_1 \vdash v_1(f) \wedge v_2(0)$

(2)  $H_2 \vdash v_2(f) \wedge v_1(0)$

Then

(3)  $H_1 \cup H_2 \vdash v_1 \sqcup v_2(f)$

Proof: (1) and (2) imply

(4)  $H_1 \cap H_2 = \varnothing \vee f = 0$

(1) and (2) imply

(5)  $H_1 \vdash v_1 \sqcup v_2(f)$

(6)  $H_2 \vdash v_1 \sqcup v_2(f)$

(4), (5) and (6) imply (3).

### 2.3.3 Rule for separate vertices

$$\frac{H \vdash v_1(f_1) \quad H \vdash v_2(f_2) \quad s(v_1) \cap s(v_2) = \varnothing}{H \vdash v_1 \sqcup v_2(f_1 + f_2)}$$

**Soundness** Assume

(1) $\quad H \vdash v_1(f_1)$

(2) $\quad H \vdash v_2(f_2)$

(3) $\quad s(v_1) \cap s(v_2) = \varnothing$

Then

(4) $\quad H \vdash v_1 \sqcup v_2(f_1 + f_2)$

Proof: (3) implies

(5) $\quad H \vdash s(v_1) \cap s(v_2)(0)$

(1), (2), (5) and the fact that $f_1 + f_2 - 0 \leq f_1 + f_2 \leq f_1 + f_2$ imply (4).

## 3 Introducing countermeasures

### 3.1 Formal foundation extended with countermeasures

We start by extending the basic formal machinery to take countermeasures into consideration.

#### 3.1.1 Timed events with countermeasures

$\mathbb{C}$ denotes the set of all countermeasures. To record treatments each timed event is extended with a possibly empty set of countermeasures. A timed event with an empty set of countermeasures is untreated, while a timed event with a nonempty set is treated by the countermeasures in the set. Hence, a timed event is from this point onwards an element of

$$\mathbb{E} \times \mathbb{T} \times \mathbb{P}(\mathbb{C})$$

#### 3.1.2 Histories with countermeasures

The notion of history is generalised straightforwardly to deal with timed events with countermeasures as follows:

$$\mathbb{H} \stackrel{\text{def}}{=} \{ \ h \in (\mathbb{E} \times \mathbb{T} \times \mathbb{P}(\mathbb{C}))^\infty \ | \\ \forall n \in \mathbb{N} : \pi_2.h[n] \leq \pi_2.h[n+1] \\ \forall t \in \mathbb{T} : \exists n \in \mathbb{N} : \pi_2.h[n] > t \quad \}$$

The truncation operator

$$\_|\_ \in \mathbb{H} \times \mathbb{T} \to (\mathbb{E} \times \mathbb{T} \times \mathbb{P}(\mathbb{C}))^*$$

is generalised accordingly.

## 3.2 Syntax extended with countermeasures

The next step is to generalize the notion of risk graph.

### 3.2.1 Risk graphs

A risk graph with treatments is a tuple of five sets $(V, C, R_l, R_e, R_d)$ where

$V \subseteq \mathbb{P}(\mathbb{E}) \times \mathbb{F}$,
$C \subseteq \mathbb{C}$,
$R_l \subseteq V \times \mathbb{R}^+ \times V$,
$R_e \subseteq C \times [0,1] \times V$,
$R_d \subseteq C \times [0,1] \times R_e$

We refer to the elements of $V$ as the set of vertices, $C$ as the set of countermeasures, and to $R_l, R_e, R_d$ as the leads-to relations, the effects relations and the dependency relations, respectively.

We use $v(f)$ to denote a vertex, $c$ to denote a countermeasure, $\xrightarrow{l}$ to denote a leads-to relation, $\xrightarrow{e}$ to denote an effects relation and $\xrightarrow{d}$ to denote a dependency relation.

### 3.2.2 Vertex expressions

The set of vertex expressions is the smallest set $X_V$ such that

$v \in \mathbb{P}(\mathbb{E}) \wedge cs \in \mathbb{P}(\mathbb{C}) \Rightarrow v_{cs} \in X_V$
$v, v' \in X_V \Rightarrow v \sqcup v' \in X_V \wedge v \sqcap\!\!| \, v' \in X_V$

We need a function

$s \in X_V \to \mathbb{P}(\mathbb{E})$

that for any vertex expression calculates its set of events. Formally, $s$ is defined recursively as follows:

$$s(v) \stackrel{\text{def}}{=} \begin{cases} v' & \text{if } v = v'_{cs} \\ s(v_1) \cup s(v_2) & \text{if } v = v_1 \sqcup v_2 \\ s(v_2) & \text{if } v = v_1 \sqcap\!\!| \, v_2 \end{cases}$$

### 3.2.3 Risk graph formula

A risk graph formula is of one of the following four forms

$$H \vdash c \xrightarrow{e}_{cs} v, \quad H \vdash c \xrightarrow{d} (c' \xrightarrow{e}_{cs} v), \quad H \vdash v'(f), \quad H \vdash v' \xrightarrow{r} v''$$

where

- $H \in \mathbb{P}(\mathbb{H})$,
- $c, c' \in \mathbb{C}$ where $c \neq c'$,

- $e, d \in [0, 1]$,
- $cs \in \mathbb{P}(\mathbb{C})$ where $c, c' \notin cs$,
- $v \in \mathbb{P}(\mathbb{E})$,
- $v', v'' \in X_V$,
- $f \in \mathbb{F}$,
- $r \in \mathbb{R}^+$.

## 3.3 Semantics extended with countermeasures

The semantics of a risk graph formula is defined recursively as before. In particular, the definitions are unchanged in the case of

$$[\![\, H \vdash v_1 \sqcup v_2(f) \,]\!], \quad [\![\, H \vdash v_1 \sqcap v_2(f) \,]\!], \quad [\![\, H \vdash v_1 \xrightarrow{r} v_2 \,]\!]$$

The vertex base-case must however be updated to take countermeasures into account:

$$[\![\, H \vdash v_{cs}(f) \,]\!] \stackrel{\text{def}}{=}$$
$$\forall h \in H :$$
$$f = \lim_{t \to \infty} \frac{\#((v \times \mathbb{T} \times \mathbb{P}(\mathbb{C} \setminus cs)) \, \textcircled{S} \, (h|_t))}{t}$$

Hence, we only take into consideration those events in $v$ that are not treated by a countermeasure in $cs$.

In the case of the effects relation the semantics is defined as follows:

$$[\![\, H \vdash c \xrightarrow{e}_{cs} v \,]\!] \stackrel{\text{def}}{=}$$
$$\exists f_1, f_2 \in \mathbb{F} :$$
$$[\![\, H \vdash v_{cs}(f_1) \,]\!]$$
$$[\![\, H \vdash v_{cs \cup \{c\}}(f_2) \,]\!]$$
$$f_1 \neq 0 \Rightarrow e = \frac{f_1 - f_2}{f_1}$$

Hence, $e$ is the fraction of $v$ events whose set of countermeasures contains $c$ but no countermeasure in $cs$.

Also the dependency relation captures a fraction:

$$[\![\, H \vdash c \xrightarrow{d} (c' \xrightarrow{e}_{cs} v) \,]\!] \stackrel{\text{def}}{=}$$
$$[\![\, H \vdash c' \xrightarrow{e}_{cs} v \,]\!] \Rightarrow$$
$$\exists e' \in [0, 1] :$$
$$[\![\, H \vdash c' \xrightarrow{e'}_{cs \cup \{c\}} v \,]\!]$$
$$e \neq 0 \Rightarrow d = 1 - \frac{e'}{e}$$

Hence, $d$ is the fraction of $v$ treated by countermeasure $c'$ that is also treated by countermeasure $c$.

## 3.4 Calculus extended with countermeasures

### 3.4.1 Rule for countermeasure effect

$$\frac{H \vdash c \xrightarrow{e}_{cs} v \quad H \vdash v_{cs}(f)}{H \vdash v_{cs \cup \{c\}}(f \cdot \overline{e})}$$

**Soundness** Assume

(1) $H \vdash c \xrightarrow{e}_{cs} v$

(2) $H \vdash v_{cs}(f)$

Then

(3) $H \vdash v_{cs \cup \{c\}}(f \cdot \overline{e})$

Proof: (1) implies there are $f_1, f_2 \in \mathbb{F}$ such that

(4) $[\![ H \vdash v_{cs}(f_1) ]\!]$

(5) $[\![ H \vdash v_{cs \cup \{c\}}(f_2) ]\!]$

(6) $f_1 \neq 0 \Rightarrow e = \frac{f_1 - f_2}{f_1}$

(2) and (4) imply

(7) $f = f_1$

There are two cases to consider:

- Assume

    (8) $f_1 = 0$

    (4) and (7) imply

    (9) $[\![ H \vdash v_{cs \cup \{c\}}(0) ]\!]$

    (7) and (8) imply

    (10) $f = 0$

    (9), (10) and $0 \cdot \overline{e} = 0$ imply (3).

- Assume

    (11) $f_1 \neq 0$

    (6), (7) and (11) imply

    (12) $e = \frac{f - f_2}{f}$

    (12) implies

    (13) $\frac{f_2}{f} = 1 - e$

    (13) implies

    (14) $f_2 = f \cdot \overline{e}$

    (5) and (14) imply (3).

### 3.4.2 Rule for countermeasure dependency

$$\frac{H \vdash c \xrightarrow{d} (c' \xrightarrow{e}_{cs} v) \quad H \vdash c' \xrightarrow{e}_{cs} v}{H \vdash c' \xrightarrow{e \cdot \overline{d}}_{cs \cup \{c\}} v}$$

**Soundness** Assume

(1) $\quad H \vdash c \xrightarrow{d} (c' \xrightarrow{e}_{cs} v)$

(2) $\quad H \vdash c' \xrightarrow{e}_{cs} v$

Then

(3) $\quad H \vdash c' \xrightarrow{e \cdot \overline{d}}_{cs \cup \{c\}} v$

Proof: There are two cases to consider:

- Assume

    (4) $\quad e \neq 0$

    (1), (2) and (4) imply there is $e' \in [0,1]$ such that

    (5) $\quad [\![ H \vdash c' \xrightarrow{e'}_{cs \cup \{c\}} v ]\!]$

    (6) $\quad d = 1 - \frac{e'}{e}$

    (6) implies

    (7) $\quad \frac{e'}{e} = 1 - d = \overline{d}$

    (5) and (7) imply (3).

- Assume

    (8) $\quad e = 0$

    (2), (8) and the semantics of the effects relation imply (3).

## 4 Introducing consequences

### 4.1 Formal foundation extended with consequences

We start by extending the basic formal machinery to take consequences into consideration.

#### 4.1.1 Timed events with consequences

$\mathbb{I}$ denotes the set of all consequences (or impacts). To facilitate arithmetic operations on consequences we assume that

$\mathbb{I} \stackrel{\text{def}}{=} \mathbb{R}^+$

To record consequences each timed event is extended with an additional component characterizing the consequence of this event with respect to the various combinations of countermeasures. A timed event is from this point onwards an element of

$$\mathbb{E} \times \mathbb{T} \times \mathbb{P}(\mathbb{C}) \times (\mathbb{P}(\mathbb{C}) \to \mathbb{I})$$

For any timed event $e$ we require

$$c \subseteq c' \Rightarrow (\pi_4.e)(c) \geq (\pi_4.e)(c')$$

Hence, adding a countermeasure will never increase the consequence.

#### 4.1.2 Histories with consequences

The notion of history is generalised straightforwardly to deal with consequences as follows:

$$\begin{aligned} \mathbb{H} \stackrel{\text{def}}{=} \{ \quad & h \in (\mathbb{E} \times \mathbb{T} \times \mathbb{P}(\mathbb{C}) \times (\mathbb{P}(\mathbb{C}) \to \mathbb{I}))^\infty \mid \\ & \forall\, n \in \mathbb{N} : \pi_2.h[n] \leq \pi_2.h[n+1] \\ & \forall\, t \in \mathbb{T} : \exists\, n \in \mathbb{N} : \pi_2.h[n] > t \qquad \} \end{aligned}$$

The truncation operator

$$\_\lfloor\_ \in \mathbb{H} \times \mathbb{T} \to (\mathbb{E} \times \mathbb{T} \times \mathbb{P}(\mathbb{C}) \times (\mathbb{P}(\mathbb{C}) \to \mathbb{I}))^*$$

is generalised accordingly.

### 4.2 Syntax extended with consequences

The next step is to generalize the notion of risk graph.

#### 4.2.1 Risk graphs

The notion of risk graph is a tuple of five sets $(V, C, R_l, R_e, R_d)$ where

$$V \subseteq \mathbb{P}(\mathbb{E}) \times \mathbb{F} \times \mathbb{I},$$
$$C \subseteq \mathbb{C},$$
$$R_l \subseteq V \times \mathbb{R}^+ \times V,$$
$$R_e \subseteq C \times [0,1] \times [0,1] \times V,$$
$$R_d \subseteq C \times [0,1] \times [0,1] \times R_e$$

We use $v(f, i)$ to denote a vertex, $\xrightarrow{(e_f, e_i)}$ to denote an effects relation and $\xrightarrow{(d_f, d_i)}$ to denote a dependency relation. The remaining conventions are as before.

#### 4.2.2 Vertex expressions

The notion of vertex expression is left unchanged.

### 4.2.3 Risk graph formula

A risk graph formula is of one of the following four forms

$$H \vdash c \xrightarrow{(e_f, e_i)}_{cs} v, \quad H \vdash c \xrightarrow{(d_f, d_i)} (c' \xrightarrow{(e_f, e_i)}_{cs} v), \quad H \vdash v'(f, i), \quad H \vdash v' \xrightarrow{r} v''$$

where

- $H \in \mathbb{P}(\mathbb{H})$,
- $c, c' \in \mathbb{C}$ where $c \neq c'$,
- $e_f, e_i, d_f, d_i \in [0, 1]$,
- $cs \in \mathbb{P}(\mathbb{C})$ where $c, c' \notin cs$,
- $v \in \mathbb{P}(\mathbb{E})$,
- $v', v'' \in X_V$,
- $f \in \mathbb{F}$,
- $i \in \mathbb{I}$,
- $r \in \mathbb{R}^+$.

## 4.3 Semantics extended with consequences

$[\![\, H \vdash v_{cs}(f, i) \,]\!] \stackrel{\text{def}}{=}$
    $\forall\, h \in H :$
        let
            $x = (v \times \mathbb{T} \times \mathbb{P}(\mathbb{C} \setminus cs) \times (\mathbb{P}(\mathbb{C}) \to \mathbb{I})) \circledS h$
        in
            $\#x = 0 \Rightarrow$
                $f = 0$
                $i = 0$
            $\#x \neq 0 \Rightarrow$
                $f = \lim_{t \to \infty} \frac{\#(x|_t)}{t}$
                $i = \lim_{t \to \infty} \frac{\sum_{1 \leq j \leq \#(x|_t)} \pi_4 \cdot x[j](cs)}{\#(x|_t)}$

$[\![\, H \vdash v_1 \sqcup v_2(f, i) \,]\!] \stackrel{\text{def}}{=}$
    $\exists\, f_1, f_2, f_3 \in \mathbb{F} :$
        $[\![\, H \vdash v_1(f_1, i) \,]\!]$
        $[\![\, H \vdash v_2(f_2, i) \,]\!]$
        $[\![\, H \vdash s(v_1) \cap s(v_2)(f_3, i) \,]\!]$
        $f_1 + f_2 - f_3 \leq f \leq f_1 + f_2$

$$[\![ H \vdash v_1 \sqcap v_2(f, i) ]\!] \stackrel{\text{def}}{=}$$
$$\exists\, r \in \mathbb{R}^+;\ f_1, f_2 \in \mathbb{F};\ i' \in \mathbb{I}:$$
$$[\![ H \vdash v_1(f_1, i') ]\!]$$
$$[\![ H \vdash v_2(f_2, i) ]\!]$$
$$f = f_1 \cdot r$$
$$f \leq f_2$$

$$[\![ H \vdash v_1 \xrightarrow{r} v_2 ]\!] \stackrel{\text{def}}{=}$$
$$\exists f_1, f_2 \in \mathbb{F};\ i_1, i_2 \in \mathbb{I}:$$
$$[\![ H \vdash v_1(f_1, i_1) ]\!]$$
$$[\![ H \vdash v_2(f_2, i_2) ]\!]$$
$$f_2 \geq f_1 \cdot r$$

$$[\![ H \vdash c \xrightarrow{(e_f, e_i)}_{cs} v ]\!] \stackrel{\text{def}}{=}$$
$$\exists f_1, f_2 \in \mathbb{F};\ i_1, i_2 \in \mathbb{I}:$$
$$[\![ H \vdash v_{cs}(f_1, i_1) ]\!]$$
$$[\![ H \vdash v_{cs \cup \{c\}}(f_2, i_2) ]\!]$$
$$f_1 \neq 0 \Rightarrow e_f = \frac{f_1 - f_2}{f_1}$$
$$i_1 \neq 0 \Rightarrow e_i = \frac{i_1 - i_2}{i_1}$$

$$[\![ H \vdash c \xrightarrow{(d_f, d_i)} (c' \xrightarrow{(e_f, e_i)}_{cs} v) ]\!] \stackrel{\text{def}}{=}$$
$$[\![ H \vdash c' \xrightarrow{(e_f, e_i)}_{cs} v ]\!] \Rightarrow$$
$$\exists\, e'_f, e'_i \in [0, 1]:$$
$$[\![ H \vdash c' \xrightarrow{(e'_f, e'_i)}_{cs \cup \{c\}} v ]\!]$$
$$e_f \neq 0 \Rightarrow d_f = 1 - \frac{e'_f}{e_f}$$
$$e_i \neq 0 \Rightarrow d_i = 1 - \frac{e'_i}{e_i}$$

## 4.4 Calculus extended with consqeuences

### 4.4.1 Rule for leads-to

$$\frac{H \vdash v_1(f_1, i_1) \quad H \vdash v_1 \xrightarrow{r} v_2 \quad H \vdash v_2(f_2, i_2)}{H \vdash v_1 \sqcap v_2(f_1 \cdot r, i_2)}$$

**Soundness** We need an additional premise to conclude that $i_2$ is the impact of $v_2$. Except for that the introduction of consequences is irrelevant for the validity of the rule. Hence, the soundness follows from the soundness of Rule 2.3.1.

### 4.4.2 Rule for mutually exclusive vertices

$$\frac{H_1 \vdash v_1(f,i) \land v_2(0,i) \quad H_2 \vdash v_2(f,i) \land v_1(0,i)}{H_1 \cup H_2 \vdash v_1 \sqcup v_2(f,i)}$$

**Soundness** The introduction of consequences is irrelevant for the validity of the rule. Hence, the soundness follows from the soundness of Rule 2.3.2.

### 4.4.3 Rule for separate vertices

$$\frac{H \vdash v_1(f_1,i) \quad H \vdash v_2(f_2,i) \quad s(v_1) \cap s(v_2) = \varnothing}{H \vdash v_1 \sqcup v_2(f_1+f_2,i)}$$

**Soundness** The introduction of consequences is irrelevant for the validity of the rule. Hence, the soundness follows from the soundness of Rule 2.3.3.

### 4.4.4 Rule for countermeasure effect

$$\frac{H \vdash c \xrightarrow{(e_f,e_i)}_{cs} v \quad H \vdash v_{cs}(f,i)}{H \vdash v_{cs \cup \{c\}}(f \cdot \overline{e_f}, i \cdot \overline{e_i})}$$

**Soundness** Assume

(1) $H \vdash c \xrightarrow{(e_f,e_i)}_{cs} v$
(2) $H \vdash v_{cs}(f,i)$

Then

(3) $H \vdash v_{cs \cup \{c\}}(f \cdot \overline{e_f}, i \cdot \overline{e_i})$

Proof: The soundness of the frequency deduction follows from the soundness of Rule 3.4.1. Hence, we focus only on the consequence deduction. (1) implies there are $f_1, f_2 \in \mathbb{F}$ and $i_1, i_2 \in \mathbb{I}$ such that

(4) $[\![ H \vdash v_{cs}(f_1,i_1) ]\!]$
(5) $[\![ H \vdash v_{cs \cup \{c\}}(f_2,i_2) ]\!]$
(6) $i_1 \neq 0 \Rightarrow e_i = \frac{i_1 - i_2}{i_1}$

(2) and (4) imply

(7) $i = i_1$

There are two cases to consider:

- Assume

    (8) $i_1 = 0$

(4), (5) and (7) imply

(9) $[\![ H \vdash v_{cs \cup \{c\}}(f_2, 0) ]\!]$

(7) and (8) imply

(10) $i = 0$

(9), (10) and $0 \cdot \overline{e_i} = 0$ imply (3).

- Assume

  (11) $i_1 \neq 0$

  (6), (7) and (11) imply

  (12) $e_i = \frac{i - i_2}{i}$

  (12) implies

  (13) $\frac{i_2}{i} = 1 - e_i$

  (13) implies

  (14) $i_2 = i \cdot \overline{e_i}$

  (5) and (14) imply (3).

### 4.4.5 Rule for countermeasure dependency

$$\frac{H \vdash c \xrightarrow{(d_f, d_i)} (c' \xrightarrow{(e_f, e_i)}_{cs} v) \quad H \vdash c' \xrightarrow{(e_f, e_i)}_{cs} v}{H \vdash c' \xrightarrow{(e_f \cdot \overline{d_f},\, e_i \cdot \overline{d_i})}_{cs \cup \{c\}} v}$$

**Soundness** Assume

(1) $H \vdash c \xrightarrow{(d_f, d_i)} (c' \xrightarrow{(e_f, e_i)}_{cs} v)$
(2) $H \vdash c' \xrightarrow{(e_f, e_i)}_{cs} v$

Then

(3) $H \vdash c' \xrightarrow{(e_f \cdot \overline{d_f},\, e_i \cdot \overline{d_i})}_{cs \cup \{c\}} v$

Proof: The soundness of the frequency deduction follows from the soundness of Rule 4.4.5. Hence, we focus only on the consequence deduction. There are two cases to consider:

- Assume

  (4) $e_i \neq 0$

(1), (2) and (4) imply there are $e'_f, e'_i \in [0,1]$ such that

(5) $[\![ H \vdash c' \xrightarrow{(e'_f, e'_i)}_{cs \cup \{c\}} v ]\!]$

(6) $d_i = 1 - \frac{e'_i}{e_i}$

(6) implies

(7) $\frac{e'_i}{e_i} = 1 - d_i = \overline{d_i}$

(5) and (7) imply (3).

- Assume

    (8) $e_i = 0$

(2), (8) and the constraint that adding a countermeasure will never increase the consequence imply (3).

# 5 Introducing intervals

## 5.1 Syntax extended with intervals

The syntax is as before with the exception that we now have intervals where we earlier had singular values.

## 5.2 Semantics extended with intervals

The semantics is generalised to intervals in a point-wise manner:

$[\![ H \vdash v_{cs}(F, I) ]\!] \stackrel{\text{def}}{=}$
$\quad \forall h \in H; \exists f \in F; i \in I:$
$\qquad [\![ \{h\} \vdash v_{cs}(f, i) ]\!]$

$[\![ H \vdash v_1 \sqcup v_2(F, I) ]\!] \stackrel{\text{def}}{=}$
$\quad \forall h \in H; \exists f \in F; i \in I:$
$\qquad [\![ \{h\} \vdash v_1 \sqcup v_2(f, i) ]\!]$

$[\![ H \vdash v_1 \sqcap v_2(F, I) ]\!] \stackrel{\text{def}}{=}$
$\quad \forall h \in H; \exists f \in F; i \in I:$
$\qquad [\![ \{h\} \vdash v_1 \sqcup v_2(f, i) ]\!]$

$[\![ H \vdash v_1 \xrightarrow{R} v_2 ]\!] \stackrel{\text{def}}{=}$
$\quad \forall h \in H; \exists r \in R:$
$\qquad [\![ \{h\} \vdash v_1 \xrightarrow{r} v_2 ]\!]$

$$[\![ H \vdash c \xrightarrow{(E_F, E_I)}_{cs} v ]\!] \stackrel{\mathsf{def}}{=}$$
$$\forall h \in H; \, \exists \, e_f \in E_F, e_i \in E_I :$$
$$[\![ \{h\} \vdash c \xrightarrow{(e_f, e_i)}_{cs} v ]\!]$$

$$[\![ H \vdash c \xrightarrow{(D_F, D_I)} (c' \xrightarrow{(E_F, E_I)}_{cs} v) ]\!] \stackrel{\mathsf{def}}{=}$$
$$\forall h \in H; \, \exists \, d_f \in D_F; \, d_i \in D_I; \, e_f \in E_F; \, e_i \in E_I :$$
$$[\![ \{h\} \vdash c \xrightarrow{(d_f, d_i)} (c' \xrightarrow{(e_f, e_i)}_{cs} v) ]\!]$$

## 5.3 Calculus extended with intervals

### 5.3.1 Rule for leads-to

$$\frac{H \vdash v_1(F_1, I_1) \quad H \vdash v_1 \xrightarrow{R} v_2 \quad H \vdash v_2(F_2, I_2)}{H \vdash v_1 \sqcap v_2(F_1 \cdot R, I_2)}$$

**Soundness** By pointwise application of Rule 4.4.1.

### 5.3.2 Rule for mutually exclusive vertices

$$\frac{H_1 \vdash v_1(F, I) \wedge v_2(\{0\}, I) \quad H_2 \vdash v_2(F, I) \wedge v_1(\{0\}, I)}{H_1 \cup H_2 \vdash v_1 \sqcup v_2(F, I)}$$

**Soundness** By pointwise application of Rule 4.4.2.

### 5.3.3 Rule for separate vertices

$$\frac{H \vdash v_1(F_1, I) \quad H \vdash v_2(F_2, I) \quad s(v_1) \cap s(v_2) = \varnothing}{H \vdash v_1 \sqcup v_2(F_1 + F_2, I)}$$

**Soundness** By pointwise application of Rule 4.4.3.

### 5.3.4 Rule for countermeasure effect

$$\frac{H \vdash c \xrightarrow{(E_F, E_I)}_{cs} v \quad H \vdash v_{cs}(F, I)}{H \vdash v_{cs \cup \{c\}}(F \cdot \overline{E_F}, I \cdot \overline{E_I})}$$

**Soundness** By pointwise application of Rule 4.4.4.

### 5.3.5 Rule for countermeasure dependency

$$\frac{H \vdash c \xrightarrow{(D_F, D_I)} (c' \xrightarrow{(E_F, E_I)}_{cs} v) \quad H \vdash c' \xrightarrow{(E_F, E_I)}_{cs} v}{H \vdash c' \xrightarrow{(E_F \cdot \overline{D_F}, E_I \cdot \overline{D_I})}_{cs \cup \{c\}} v}$$

**Soundness** By pointwise application of Rule 4.4.5.

### 5.3.6 Rule for arbitrary vertices

$$\frac{H \vdash v_1(F_1, I) \quad H \vdash v_2(F_2, I)}{H \vdash v_1 \sqcup v_2([\max(\{\min(F_1), \min(F_2)\}), \max(F_1) + \max(F_2)], I)}$$

**Soundness** The upper bound corresponds to the case where the set of events of the two vertices in a history are disjoint, while the lower bound corresponds to the case where the set of events in a history belonging to one of the vertices is fully contained in the history's set of events belonging to the other vertex.

# 6 Relating CORAS to Risk Graphs

We distinguish between two kinds of CORAS elements, namely the set $\mathbb{E}_{UI}$ of unwanted elements, and the set $\mathbb{E}_{TS}$ of scenario elements. We assume that

$$\mathbb{E}_{UI} \cap \mathbb{E}_{TS} = \varnothing$$

We refer to the sequences in $\mathbb{E}_{TS}{}^*$ as the threat scenarios elements. An unwanted incident in CORAS may be thought of as a set of unwanted elements, while a threat scenario corresponds to a set of threat scenario elements.

A timed CORAS event is a quadruple of the following type

$$(\mathbb{E}_{UI} \cup \mathbb{E}_{TS}{}^*) \times \mathbb{T} \times \mathbb{P}(\mathbb{C}) \times (\mathbb{P}(\mathbb{C}) \to \mathbb{I})$$

While an unwanted incident element is instantanious a threat scenario element is not. The timestamp of a threat scenario element denotes its time of termination. In CORAS only unwanted incidents may have a consequence. Hence, in the case of threat scenario elements, any set of countermeasures is mapped to 0.

The relationship between a timed CORAS event and a timed risk graph even is defined by a function *map* such that

$$map(e, t, co, im) \stackrel{\text{def}}{=} (m(e), t, co, im)$$

where

$$m \in \mathbb{E}_{UI} \cup \mathbb{E}_{TS}{}^* \to \mathbb{E}$$

is a bijective function. This means that

$$m(e) = m(e') \Rightarrow e = e'$$